\documentclass[aps,prx,twocolumn,amsmath,amssymb,footinbib,showpacs,longbibliography,superscriptaddress]{revtex4-1}

\usepackage{times}
\usepackage{ulem}

\usepackage[english]{babel}
\usepackage{latexsym}
\usepackage{graphics}
\usepackage{subfigure}
\usepackage{epsfig}
\usepackage{color}
\usepackage{hyperref}
\usepackage{braket} 
\usepackage[T1]{fontenc}
\usepackage[latin9]{inputenc}
\usepackage{amstext}
\usepackage{amssymb}
\usepackage{graphicx}
\usepackage{esint}
\usepackage{braket}
\usepackage{babel}
\usepackage{amsmath}
\usepackage{float}

\hypersetup{
colorlinks=true,
citecolor=blue,
linkcolor=red,
urlcolor=black
}


\newcommand{\think}[1]{{\color{red} {#1}}}

\newcommand{\todo}[1]{{\color{blue} {#1}}}
\definecolor{purple}{rgb}{1,0,1}

\newcommand{\ie}{i.e.}

\newcommand{\Er}{E_{\textrm{r}}}

\newcommand{\Ek}{E_{\textrm{k}}}
\newcommand{\Vc}{V_{\textrm{c}}}
\newcommand{\Jc}{J_{\textrm{c}}}
\newcommand{\oneD}{\textrm{1D}}
\newcommand{\twoD}{\textrm{2D}}

\newcommand{\kB}{k_{\textrm{\tiny {B}}}}

\newcommand{\fc}{f_{\textrm{c}}}

\begin{document}

\title{
Universal scaling of many-body effects in quantum tunneling
}

\affiliation{State Key Laboratory of Photonics and Communications, School of Electronics, Peking University, Beijing 100871, China}
\affiliation{DQMP, University of Geneva, 24 Quai Ernest-Ansermet, CH-1211 Geneva, Switzerland}
\affiliation{Institute of Carbon-based Thin Film Electronics, Peking University, Shanxi, Taiyuan 030012, China}

\author{Hongmian Shui}
\homepage{These authors contributed equally to this work.}
\affiliation{State Key Laboratory of Photonics and Communications, School of Electronics, Peking University, Beijing 100871, China}
\affiliation{Institute of Carbon-based Thin Film Electronics, Peking University, Shanxi, Taiyuan 030012, China}
\author{Chi-Kin Lai}
\homepage{These authors contributed equally to this work.}
\author{Chengyang Wu}
\homepage{These authors contributed equally to this work.}
\affiliation{State Key Laboratory of Photonics and Communications, School of Electronics, Peking University, Beijing 100871, China}
\author{Lorenzo Pizzino}
\affiliation{DQMP, University of Geneva, 24 Quai Ernest-Ansermet, CH-1211 Geneva, Switzerland}
\author{Chi~Zhang}
\affiliation{State Key Laboratory of Photonics and Communications, School of Electronics, Peking University, Beijing 100871, China}
\author{Guohao Shen}
\affiliation{Institute of Carbon-based Thin Film Electronics, Peking University, Shanxi, Taiyuan 030012, China}
\author{Thierry Giamarchi}
\email{thierry.giamarchi@unige.ch}
\affiliation{DQMP, University of Geneva, 24 Quai Ernest-Ansermet, CH-1211 Geneva, Switzerland}
\author{Hepeng Yao}
\email{hepeng.yao@pku.edu.cn}
\affiliation{State Key Laboratory of Photonics and Communications, School of Electronics, Peking University, Beijing 100871, China}
\author{Xiaoji Zhou}
\email{xjzhou@pku.edu.cn}
\affiliation{State Key Laboratory of Photonics and Communications, School of Electronics, Peking University, Beijing 100871, China}
\affiliation{Institute of Carbon-based Thin Film Electronics, Peking University, Shanxi, Taiyuan 030012, China}

\date{\today}

\think{
\begin{abstract}

Quantum tunneling is fundamental to diverse phenomena and underpins a wide range of modern technologies. In the study of superconducting quantum computation and high-temperature superconducting materials, tunneling on multi-particle scale is central. Recently, several cold atom experiments successfully simulated the tunneling process in a many-particle ensemble. However, the many-body nature remains largely unexplored.  Here, we observe the universal scaling  of many-body effects in quantum tunneling process, using a hexagonal-triangular quantum simulator with independent control of barrier, temperature and interaction. In the weak-interaction regime, the critical tunneling coefficient scales parabolically with temperature under various conditions, in contrast to the linear scaling of single-particle tunneling. By further increasing the interactions beyond the mean-field regime, the scaling exponent decreases, consistent with quantum field theory predictions. Our results address the
fundamental question of how many-body effects renormalize quantum tunneling, with direct implications for correlated quantum matter and devices.


\end{abstract}
}

\maketitle


\begin{figure*}
    \centering
    \includegraphics[width=0.9\linewidth]{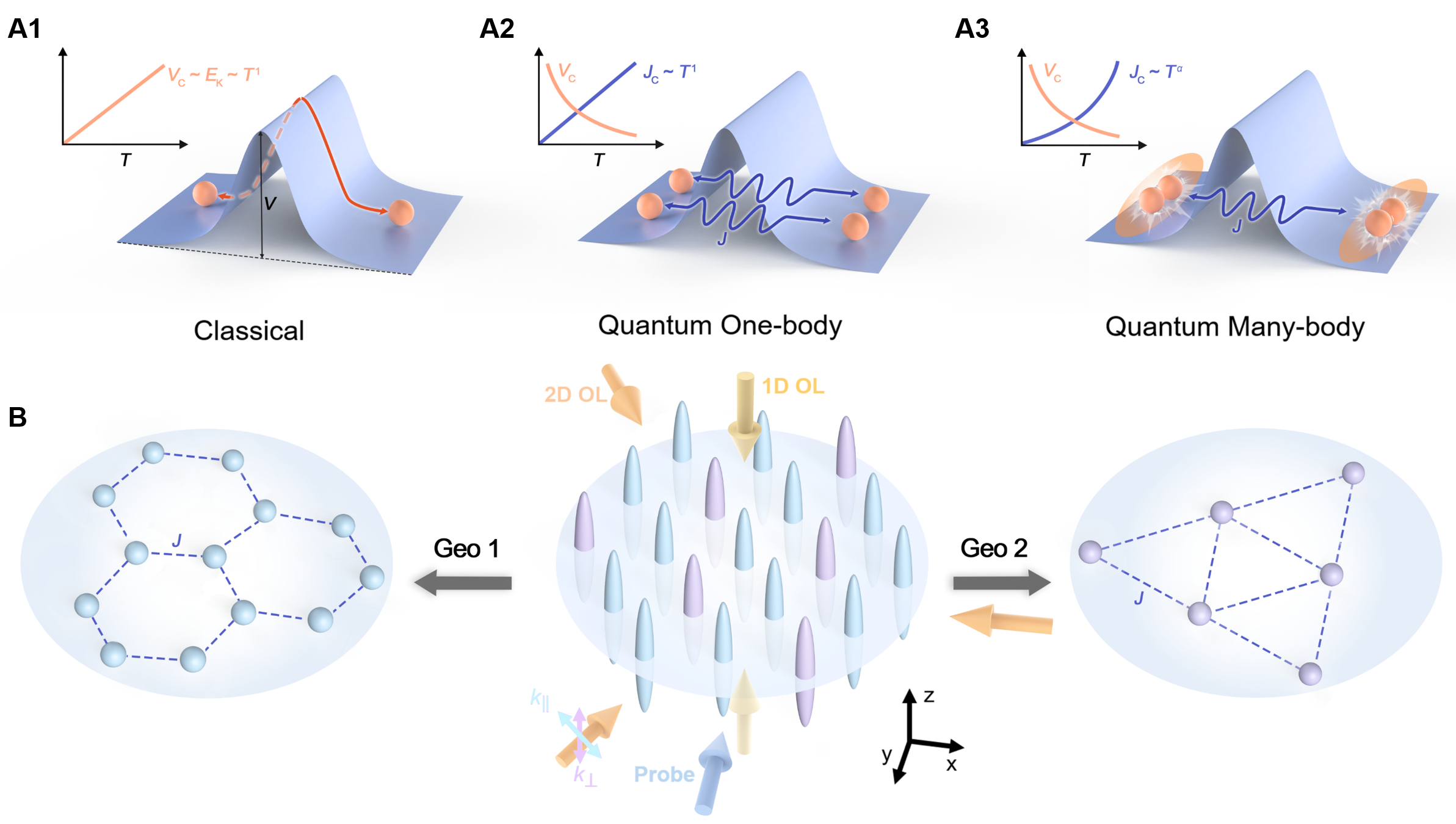}
    \caption{\textbf{Illustration of the tunneling models and experimental set-up.} $\textbf{(A1-A3)}$  Tunneling models of classical systems, quantum one-body systems, and quantum many-body systems, respectively. In classical systems $\textbf{(A1)}$, the ability of an atom to cross a  barrier is determined by its thermal motion, as shown by the red arrow. An atom with higher temperature (kinetic energy) $T\simeq E_k$  requires a sufficiently high barrier $\Vc$ to prevent it from crossing, resulting in  $\Vc\sim \Ek \sim T$. In quantum systems, the ability of an atom to cross a  barrier is determined by quantum tunneling strength $J$, as shown by the blue arrows in $\textbf{(A2)}$ and $\textbf{(A3)}$. The critical tunneling strength $\Jc$ increases with $T$, while $V_c$ decreases. In the quantum one-body case $\textbf{(A2)}$, $J$ is only related to $T$, exhibiting a linear relationship  $\Jc \sim T$. In the quantum many-body case $\textbf{(A3)}$, interaction plays a crucial role, which results in a different power-law relation $\Jc\sim T^\alpha$.   
    $\textbf{(B)}$ The experimental set-up. 
    As shown in the middle panel, three laser beams (orange arrows) at  $120^{\circ}$ intervals form a 2D traveling wave optical lattice in the $xy$-plane. 
    The lattice geometry is hexagonal or triangular when the polarization is parallel (light blue, $k_{\parallel}$) or perpendicular (purple, $k_{\perp}$) to the $xy$-plane, see left and right panels. Besides, laser beams (yellow arrows) along the $z$-axis forms a 1D standing wave optical lattice in the $z$-direction, and a probe beam (blue arrow) along the $y$-axis is used for momentum distribution measurement by absorption imaging. 
    }
    \label{fig:sch}
\end{figure*}



Quantum tunneling process is a cornerstone of quantum mechanics, acting as the fundamental mechanism for numerous natural processes and a vital bridge between basic science and cutting-edge technology. 
On a cosmic scale, it underpins stellar nuclear fusion \cite{quantum_tunneling_nuclear_fusion-1998} and Big Bang nucleosynthesis \cite{quantum_tunneling_BBN-2016},
by describing how protons and neutrons conquer the Coulomb barrier and enter the fusion channel.
In the biological realm, it facilitates essential processes such as enzyme catalysis \cite{tunneling_enzyme_catalysis-CR-2014} and DNA spontaneous mutation  \cite{tunneling_bio_DNA-1963}, where tunneling through chemical bonds and solvent molecules is crucial. 
Beyond natural phenomena, quantum tunneling principles are the operational bedrock of modern devices, including scanning tunneling microscopes \cite{STM-1982} and flash memory chips \cite{Flash_memory-2003}. In these devices, the essence is single-electron tunneling through vacuum or insulating layer.

With the advancement of superconducting quantum computing ~\cite{superconducting_computing_nature_1999,superconducting_computing_nature-2019,superconducting_computing_science_google-2020,superconducting_computing_nature_pan-2023,superconducting_computing_nature_yu-2023} and high-temperature superconducting materials ~\cite{superconducting_Cu-1986,superconducting_high_pressure-2015,superconducting_twist_graphene-2018,superconducting_lan-2019,superconducting_Ni-2019,superconducting_Ni_xue-2023},
complex tunneling processes with large particle numbers has become a key research focus.
A prime example is the macroscopic quantum effect in Josephson junctions \cite{PhysRevLett.55.1543}, where Cooper pairs undergo macroscopic tunneling. This phenomenon underpins the superconducting quantum computing \cite{superconducting_computing_nature_1999}. To probe the underlying mechanisms, several groups have simulated Josephson junction using ultracold atomic systems, 
successfully observing the tunneling dynamics via a point contact~\cite{strongly_correlated_Esslinger_science-2015} or weak barrier~\cite{Josephson-effect-BEC-BCS-Roati-science-2015}, as well as the hallmark Shapiro step \cite{doi:10.1126/science.ads8885, doi:10.1126/science.ads9061}.
Unlike traditional solid-state devices, these platforms allow direct observation of tunneling in a many-particle ensemble. 

In most multi-particle settings, the inter-particle interaction are usually non-negligible and can induce collective tunneling.
One way to characterize such process is the scaling between the critical barrier depth and temperature, see illustration in Fig.~\ref{fig:sch}A.
Considering a simple model of a particle attempting to pass a potential barrier. 
In the classical regime, the particle expends its kinetic energy $E_k$ to go over the barrier (Fig.~\ref{fig:sch}A1). By equi-partition theory, the critical barrier depth $\Vc$ scales as $\Vc\sim E_k \sim T$. Under the quantum description, a single particle can tunnel through the barrier even when $E_k<V$. This introduces a new energy scale, the tunneling coefficient $J$, which decreases with the potential depth $V$. Under this circumstance, the temperature opposes tunneling, yielding the critical scaling $\Jc\sim T$ and $V_c$ decreases with $T$ (Fig.~\ref{fig:sch}A2). Extending to the many-body case, although the competition between $J$ and $T$ remains, the scaling changes to $\Jc\sim T^{\alpha(\gamma)}$, where the exponent $\alpha\neq1$ depends on the interaction strength $\gamma$ (Fig.~\ref{fig:sch}A3). 
Understanding how interaction affects this scaling exponent is crucial, with one direct application being organic superconductors~\cite{jerome_organic_review, bergk2011superconducting, PhysRevB.71.075104}.
Although this relationship has been studied theoretically using quantum field theory~\cite{cazalilla-coupled1D-2006,Field-theory-Lorenzo-PRR-2025,Field-theory-Lorenzo-PRBL-2026}, 
experimental verification has been lacking because of the difficulty in tuning physical parameters independently, leaving the validity range of the predictions unclear.

Here, we report a cold atom realization of quantum many-body tunneling process and probe its universal nature. Unlike the single-particle case where $\Jc\sim T$, we observe a parabolic scaling $\Jc\sim T^{\alpha}$ with $\alpha=2$, matching the mean-field prediction. Using our hexagonal-triangular quantum simulator, we test the universality of this scaling law under different conditions of atom numbers, coupling structures and longitudinal properties. 
Also, by tuning the equivalent interaction strength via the longitudinal lattice, we find that the exponent $\alpha$ deviates from the meanfield value but still remains consistent with quantum field theory. Further enhancing the longitudinal lattice depth, the situation continuously reduces to the single-particle tunneling, where $\alpha$ reduces to $1$. Together, these results provide the first observation of the universal temperature scaling law for quantum many-body tunneling process in a controlled quantum ensemble. Our observations address the central question of how many-body effects renormalize quantum tunneling, and reveal the interplay of temperature, interaction, and potential barrier.


Our experimental procedure begins  with the preparation of  $^{87}\textrm{Rb}$  BEC in the $\ket{\textrm{F} = 1, m_\textrm{F} = -1}$ state with tunable atom numbers $N$ and temperatures $T$. These two parameters can be changed independently by modifying the magneto-optical trap (MOT) loading time and the final optical trap intensity during the evaporation cooling stage \cite{tian2025probinguniversalphasediagram}(see details in Supplemental material). The typical range of $N$ is from $1\times10^5$ to $3\times10^5$ while the typical temperature range is from 20 nK to 150 nK.  Following this,  the obtained BEC is adiabatically loaded into a three-dimensional (3D) optical lattice by ramping up the intensity of lattice laser beams with a rising time of 80 ms and 20 ms holding. In the middle panel of Fig.~\ref{fig:sch}B, we draw the sketch of our setup. The 3D lattice consists of a 2D traveling wave optical lattice in the $xy$-plane (2D OL, orange arrows) and an orthogonal 1D standing wave optical lattice along the $z$-axis (1D OL, yellow arrows), formed by $\lambda=1064$ nm laser beams.  It leads to lattice constants $a_{1}=\lambda/2=532$ nm in the $z$-direction and $a_{2}=2\lambda/3=709$ nm in the $xy$-plane.   

By changing the polarization of the 2D lattice beams, be it parallel (light blue, $k_{\parallel}$) or perpendicular (purple, $k_{\perp}$)  to the $xy$-plane, the 2D lattice geometry can be hexagonal (Geo1) and triangular (Geo2) correspondingly, see left and right panels of Fig.~\ref{fig:sch}B. The tunneling of adjacent lattice grids can be easily controlled by tuning the 2D lattice depth $V_{\twoD}$ in units of $\Er = 2\pi^2\hbar^2/(m\lambda^2)$, where $\Er$ is the recoil energy and  $m$ is the mass of $^{87}\textrm{Rb}$ atom. In addition, the lattice geometry also significantly affects the effect of unit well depth on tunneling strength, due to the different band structures \cite{Becker_2010}.
With such a design, we construct 1D many-body tubes with controlled tunneling.
After loading atoms into the lattice at the controlled atom number $N$ and temperature $T$, we apply a time-of-flight (TOF) process with duration of 28.5 ms and measure its momentum distribution using absorption imaging along the $y$-axis (blue arrow in Fig.~\ref{fig:sch}B). 

\begin{figure}
    \centering
    \includegraphics[width=1\linewidth]{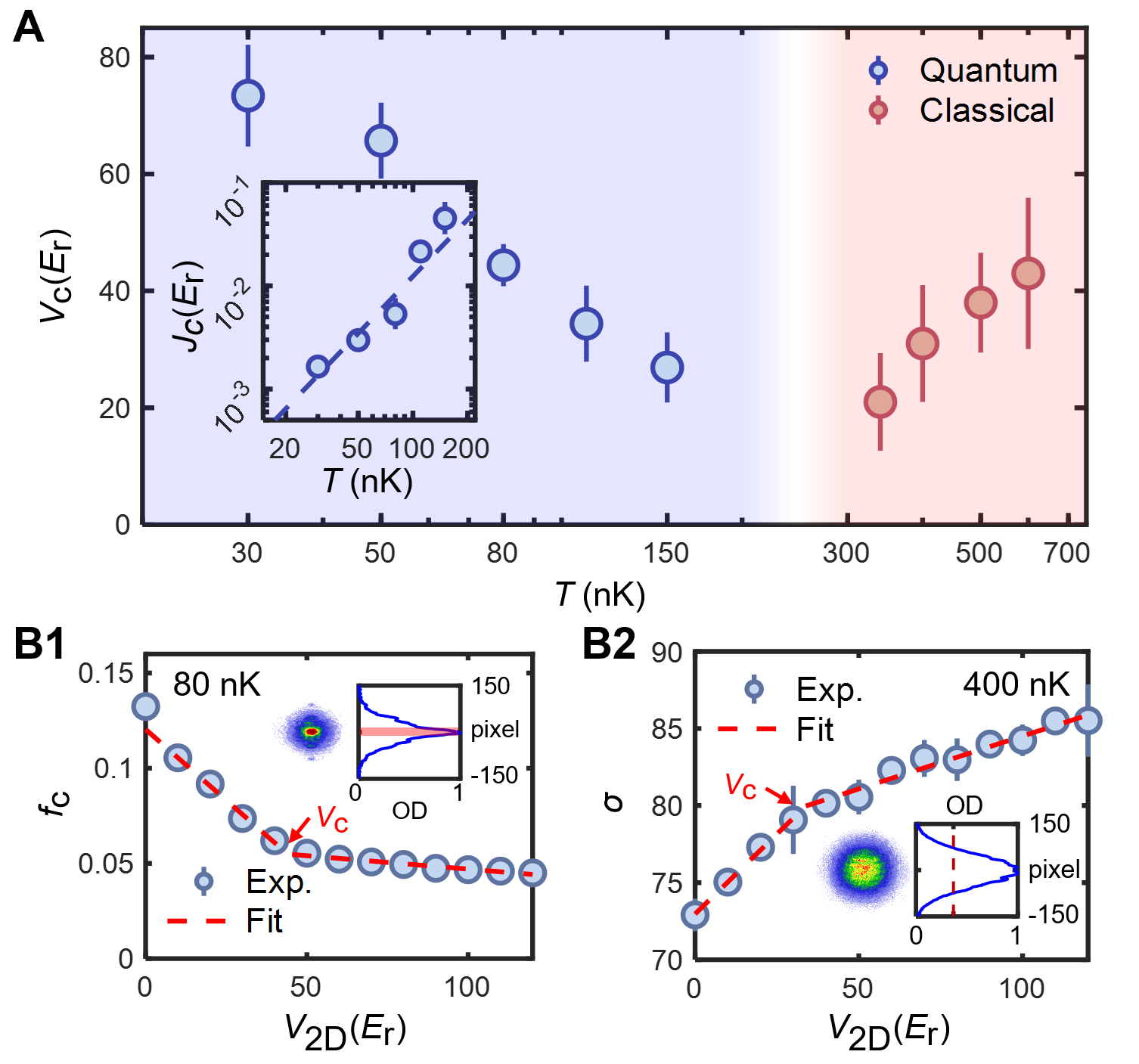}
    \caption{\textbf{The measurement of critical lattice depth $\Vc$ versus temperature $T$.} 
    $\textbf{(A)}$  The full measurement of $\Vc$ versus $T$ for fixed  atom number $N=3\times10^5$(quantum regime) and $N=1\times10^6$(classical regime) in a  hexagonal lattice. The blue and red circles represent experimental data of quantum and classical regimes, respectively. In the inner log-log plot, blue circles show the corresponding tunneling strength $\Jc$ versus $T$ of quantum regime and the blue dashed line shows the linear fitting result. 
    $\textbf{(B1-B2)}$ 
    Typical examples of determining $\Vc$ in the quantum (B1) and classical (B2) regimes. In both plots, the blue circles represent the experimental data, and the red dashed lines are the piecewise linear fits used to determine $\Vc$. The inset of (B1) shows the TOF image for the $V_{\twoD}=$ 30 $\Er$ data where the red shaded area indicates the zero momentum region.  And the inset of (B2) is the TOF image for the $V_{\twoD}=$ 10 $\Er$ data where the intersection of the blue curve and the dark red dashed line($\textrm{OD}=1/e$) represents the Gaussian radius $\sigma$. Error bars in (A) are obtained from the piecewise fit, while those in (B1-B2) represent the standard deviation of five measurements.}
    \label{fig:V_T}
\end{figure}

We start with investigating the tunneling law $\Vc$ versus $T$ in both classical and quantum regimes, correspondingly Fig.~\ref{fig:sch}A1 and A3.
In our experimental setup, the BEC phase transition temperature of the 3D system is about $T_{\textrm{BEC}}= 250$ nK. 
In Fig.~\ref{fig:V_T}A, we show the measured critical lattice depth $\Vc$ as a function of temperature $T$ in the regimes both below and above $T_{\textrm{BEC}}$ with $V_\oneD=0 \ \Er$. For $T< T_{\textrm{BEC}}$, the system is in quantum regime and the critical lattice depth can be decided by measuring the zero momentum fraction ~\cite{plisson2011coherence,guo2024observation,yao2023strongly,pizzino2025finite,tian2025probinguniversalphasediagram}
\begin{equation}
\fc=\frac{\int_{-\Delta k_x}^{+\Delta k_x}\int_{-\infty}^{+\infty}n(k_x,k_z)\mathrm{d}k_x\mathrm{d}k_z}{\int_{-\infty}^{+\infty}\int_{-\infty}^{+\infty}n(k_x,k_z)\mathrm{d}k_x\mathrm{d}k_z}
\end{equation}
with $\Delta k_x=2\pi/L_x$ and $L_x$ the system size along $z$ direction. In the inset of Fig.~\ref{fig:V_T}B1, we show a typical example of the TOF pattern for the case $T=80$ nK and $V_{\twoD}= 30\  \Er$, where $f_c$ is decided by the portion of the red shaded area. 
The corresponded $\fc -V_\twoD$ curve is shown in Fig.~\ref{fig:V_T}B1. Similarly to Refs.\cite{tian2025probinguniversalphasediagram}, we apply a piecewise fitting (red dashed line) and decide the critical point $\Vc$. Repeating this procedure for different temperatures from $30$ to $150$ nK with fixed atom number $N=3\times10^5$, we obtain the $\Vc$ data in the quantum regime (blue circles) of Fig.~\ref{fig:V_T}A. Clearly, it decreases as a function of $T$. By converting $V_c$ into an effective tunneling strength $\Jc$ according to the band structure~\cite{struck2011}, we find that the $\Jc-T$ curve exhibits a linear behavior in a log-log scale (the inset of Fig.~\ref{fig:V_T}A). It gives an algebraic scaling $\Jc\sim T^{\alpha}$ with $\alpha=1.89(26)$, which is in agreement with the physical picture in Fig.~\ref{fig:sch}A3 (see detailed discussions about the physics below).

Then we turn to the classical regime where $T\textgreater T
_{\mathrm{BEC}}$. A typical TOF image for $T=400$ nK and $ V_{\twoD}=10 \ \Er$ is shown in 
Fig.~\ref{fig:V_T}B2. Since there is no longer quantum coherence in the system, the zero momentum fraction $f_c$ is not a good criteria for the existence of barrier crossing between lattice grids. Thus, we further extract the Gaussian radius
$\sigma$ of the momentum distribution. For this observable, a piecewise fitting is still applicable from which we decide a critical $\Vc$, see Fig.~\ref{fig:V_T}B2. We repeat the same procedure for different temperatures from $340$ to $600$ nK with fixed particle number $N=1\times10^6$. The obtained $\Vc$ are plotted as red circles in Fig.~\ref{fig:V_T}A. As predicted in Fig.~\ref{fig:sch}A1, $\Vc$ increases linearly with temperature $T$ in the classical regime thanks to the equipartition theorem. By applying a linear fitting, we find the Pearson coefficient $P=0.994$.

\begin{figure*}
    \centering
    \includegraphics[width=0.9\linewidth]{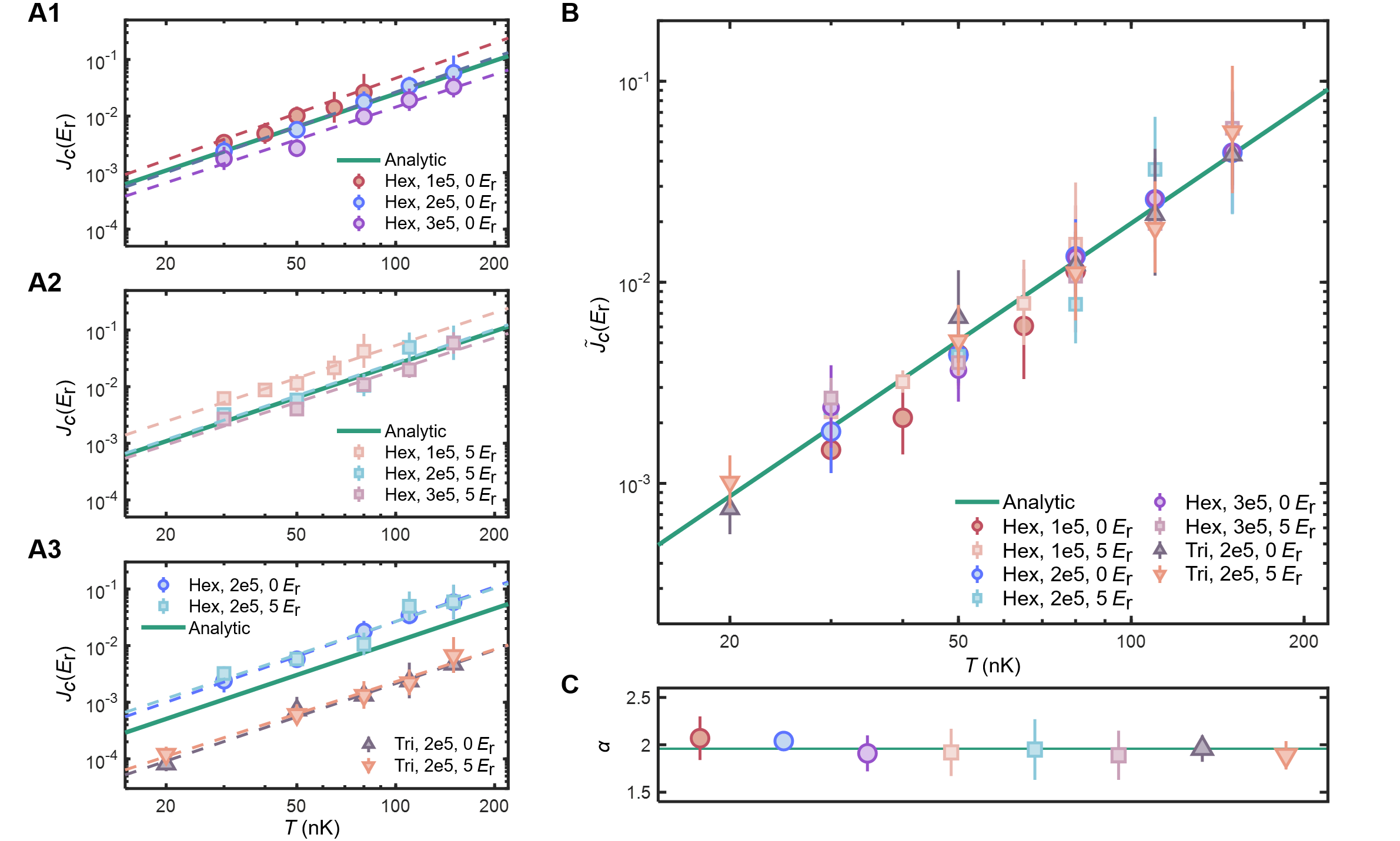}
    \caption{\textbf{The universality of the many-body tunneling scaling law.} $\textbf{(A)}$ Critical tunneling strength $\Jc$ as a function of temperature $T$. $\textbf{(A1)}$ Experimental result of different $N$ in a hexagonal lattice at $V_{\oneD} =  0\  \Er$.
    The circles with different colors represent $N=1\times 10^5$ (red circles), $2\times 10^5$ (blue circles) and $3\times 10^5$ (purple circles) results with $V_{\oneD}=0\ \Er$, respectively. The green solid line is a guide to the eye with the slope given by the theoretical prediction, $\alpha(K=14)=1.96$, and is also shown in $\textbf{(A2)}$ and $\textbf{(A3)}$.
    $\textbf{(A2)}$ Results for $V_{\oneD}=5\ \Er$ with $N=1\times10^5$ (light orange squares), $2\times10^5$ (light cyan squares), and $3\times10^5$ (dusty pink squares) for the same lattice geometry.
    $\textbf{(A3)}$ Results for different lattice geometries with $N=2\times10^5$. The gray upright triangles (orange inverted triangles) represent $V_{\oneD}=0\ \Er$ ($5\ \Er$) results in triangular lattices, while the blue circles (light cyan squares) represent  $V_{\oneD}=0\ \Er$ ($5\ \Er$) results in  hexagonal lattices. The dashed lines in $\textbf{(A)}$ with corresponding color represent the linear fitting results. 
    $\textbf{(B)}$ Rescaled $\tilde{\Jc}$ as a function of $T$ at different experimental conditions. 
    The average of all the experimental data point at $T=80$ nK is taken as the reference point. A green solid line with $\alpha(K=14)= 1.96$ crossing the reference point is plotted as a guide line. The rescaled data $\tilde{\Jc}$ are obtained by applying appropriate offsets to each dataset such that their fitted lines intersect the reference point.
    $\textbf{(C)}$ The extracted scaling exponent $\alpha$ for different atom number $N$, 1D lattice depth $V_{\oneD}$ and lattice geometries. Each data point with different colors and shapes corresponds to the slope of the linear fitting for the corresponding data in (A1-A3), and the green line represents the theoretically prediction by Eq.~\ref{eq3}.  
    Error bars in (A) and (B) are obtained from the piecewise fit, while those in (C) are obtained from the linear fit.
    }
    \label{fig:slope}
\end{figure*}

Now, we turn our focus back to the quantum regime and investigate the scaling behavior $J\sim T^{\alpha}$ in details. Near the transition point, our system can be described by the Hamiltonian of coupled-1D systems, which writes~\cite{PhysRevLett.92.130405,PhysRevB.102.195145,tian2025probinguniversalphasediagram}
\begin{align}
    \hat{H}=\sum_i\left[\hat{H}_{\textrm{1D},i}-J(\hat{a}^\dagger_{i+1}\hat{a}_{i}+\rm h.c.)\right],
    \label{eq1}
\end{align}
with $\hat{H}_{\rm 1D}$ the Hamiltonian of the 1D tubes along $z$ axis and $J$ the tunneling between these tubes.
The first term is the Lieb-Liniger model which writes
\begin{align}
    \hat{H}_{\rm 1D}=\sum_j \left[-\frac{\hbar^2}{2m}\frac{\partial^2}{\partial z_j^2}+V(z_j)\right]+\sum_{j<k}g_{\rm1D}\delta(z_j-z_k),
   \label{eq2}
\end{align}
with $g_{\rm 1D}$ the 1D coupling constant and $V(z_j)=V_{\rm1D}\cos^2{k_Lz_j}+m\omega_z^2z_j^2/2$ the external 1D lattice and harmonic potentials.
For this model, the interaction strength is governed by the Lieb-Linger parameter $\gamma=mg_{\rm 1D}/\hbar^2n$, where $n$ is the density of the 1D tube~\cite{bloch-review-2008,cazalilla2011}. The systems is in weak (strong resp.) interaction regime when $\gamma\ll 1$ ($\gamma \gg 1$ resp.). The parameter $\gamma$ is directly connected to Tomonaga-Luttinger liquid (TLL) physics through the Luttinger parameter $K(\gamma)$~\cite{haldane1980,cazalilla2011}. 
In Ref.~\cite{cazalilla-coupled1D-2006}, using a mean field approximation of the interchain tunneling and a TLL description of the tubes, Cazalilla et al. obtained an estimate of the critical tunneling strength $\Jc$. It reads
\begin{align}
    \Jc \propto T^{\frac{4K-1}{2K}},
    \label{eq3}
\end{align}
with the scaling exponent $\alpha=(4K-1)/2K$. In our experiment, the typical value of $K$ ranges from $10$ to $18$, deeply in the mean-field regime $K\gg1$. This leads to the exponent $\alpha\simeq 2$, \ie\ a square dependence of temperature.  Notably, the prefactor of the scaling in Eq.~\ref{eq3} depends on the  particle density $n$ and lattice geometry~\cite{cazalilla-coupled1D-2006}, but is independent of $T$. It therefore appears as a constant offset in a log--log plot and does not affect the scaling. 
The underlying mechanism of this parabolic dependence stems from the one-body Green function being essentially constant over the space-imaginary time regime and bounded by two limits, each on the order of the inverse temperature.
The critical tunneling, determined by the corresponding inverse susceptibility (integral of the one-body Green function over the full regime), therefore inherits the product of these bounds, giving rise to the observed $T^2$ behavior.


To verify the universality of the predicted scaling behavior under different experimental conditions, we conduct experiments with different particle numbers $N$, longitudinal constrain $V_{\oneD}$, and transverse 2D lattice geometries. 
First,  
by adapting the parameters of optical sequence as mentioned above, we change particle numbers for fixed temperatures ranging from $T=20$ nK to $200$ nK.
In practice, we measure $\Vc$ versus $T$ at $V_{\oneD} = 0 \ \Er$  with $N = 1\times 10^5$, $2\times 10^5$, and $3\times 10^5$. By converting $\Vc$ into the critical tunneling $\Jc$, we plot the $\Jc-T$ data in log-log scale, see Fig.~\ref{fig:slope}A1. The dashed lines are the corresponded linear fits, where we obtain the slopes $\alpha=2.07(23), 2.04(7), 1.91(19)$ respectively. 
For better visualization, we add a guide line with the slope of theoretical prediction $\alpha(K= 14) =1.96$ from Eq.~\ref{eq3} (solid green line). 

Then, we study the influence of the 1D optical lattice to test the applicability of the field theory in both continuous and discrete systems.  We repeat the measurements at $V_{\oneD} = 5 \ \Er$  with the same particle numbers. Similarly, we convert the results into $\Jc-T$ curves in log-log scale. The results are shown in Fig.~\ref{fig:slope}A2.  The dashed lines of the corresponding colors represent the slopes of linear fitting $\alpha= 1.92(25), 1.95(32), 1.89(26)$. Clearly, the scaling holds for the 1D discrete model and fits with the parabolic prediction from Eq.~\ref{eq3}.

Furthermore, by changing the polarization of the  2D lattice beams, we can switch between hexagonal and triangular lattice structures, which allows us to test the generality of the scaling under different lattice geometries (see details in Supplemental materials). 
We repeat the $\Vc-T$ measurements in the triangular lattice at $V_{\oneD} = 0 \ \Er$  and $5 \ \Er$ with fixed atom number $N = 2\times 10^5$.  The results are shown as gray and orange triangles in Fig.~\ref{fig:slope}A3. The dashed lines of corresponding colors represent the linear fits, whose slopes are 1.96(14) and 1.89(15). For better comparison, the results of hexagonal lattice under the same conditions are plotted. Similarly, as discussed in Ref.~\cite{cazalilla-coupled1D-2006}, the lattice geometry only influences the coordination number $z_c$ which appears directly as a prefactor in Eq.~\ref{eq3}. Therefore, in Fig.~\ref{fig:slope}A3, while the obtained slopes for the triangular and hexagonal lattices are similar, there is a clear shift between the respected groups of data in the log-log scale.



To better view the 
universality of the probed scaling law for quantum many-body tunneling, we combine all the measured data in one single plot, see Fig.~\ref{fig:slope}B. To guide the eyes, we take the average of all the data points at $T=80$ nK as reference and draw the guiding line (green solid line) with slope $\alpha(K= 14) =1.96$. In parallel, an appropriate offset is added to each group of data. 
Clearly, all the rescaled data $\tilde{\Jc}$ collapses to the guide line, which directly shows the universal scaling of all the measurements under different conditions. 
In addition, we further plots the value of the extracted scaling exponent $\alpha$ for different dataset in Fig.~\ref{fig:slope}C. The green solid line are theoretical prediction $\alpha(K=14)=1.96$.
All the experimental data fits with the theoretical prediction within $6\%$, which further confirms the parabolic scaling $\Jc\sim T^2$ and proves the field theory prediction.

\begin{figure}
    \centering
    \includegraphics[width=1\linewidth]{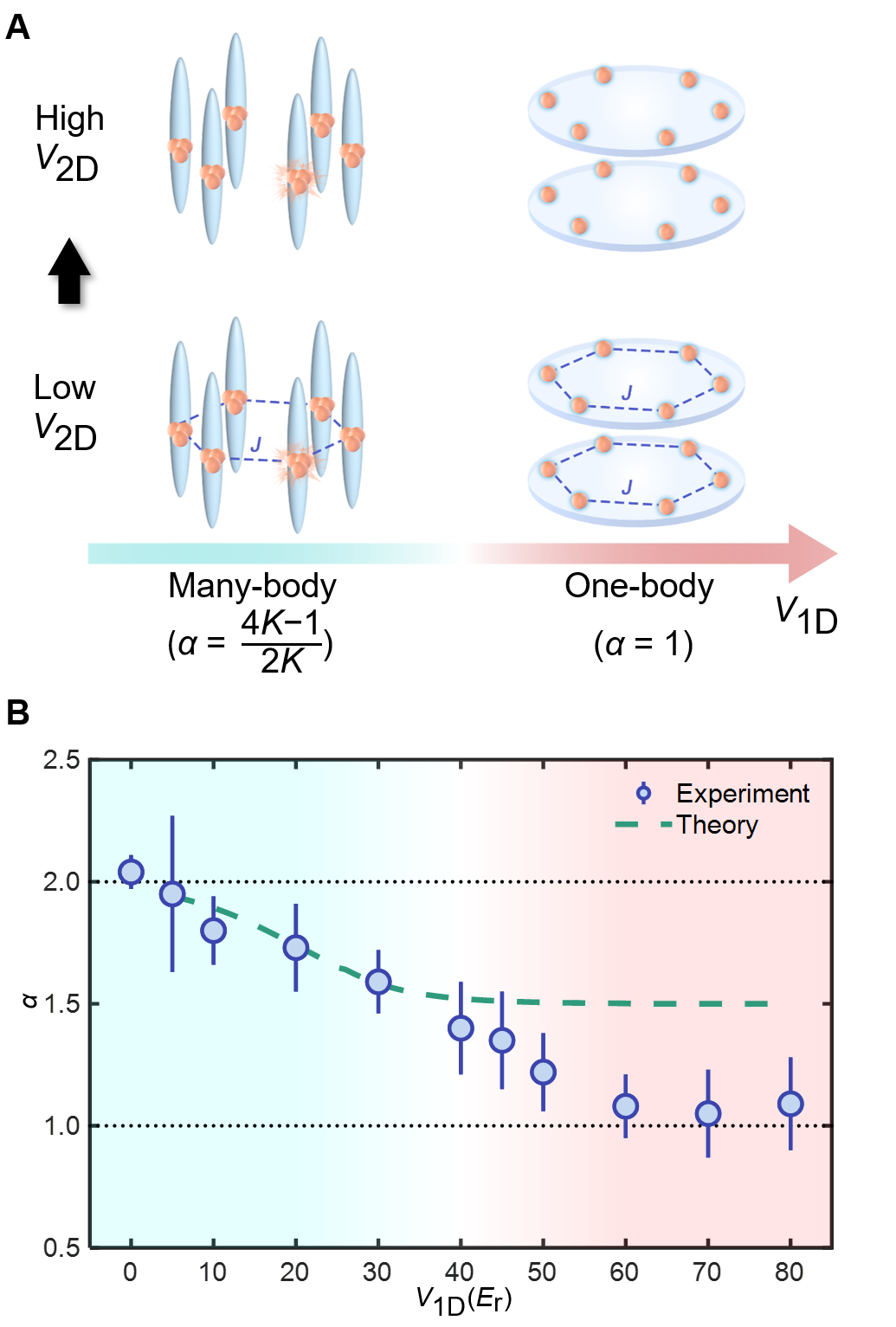}
    \caption{
    \textbf{The effect of interaction and the 
    reduction to quantum one-body tunneling.} 
    $\textbf{(A)}$ Schematic diagram of crossover process in different longitudinal lattice depth $V_{\oneD}$. The arrow with color shifting from blue to red indicates system undergoes crossover from many-body tunneling to one-body tunneling process, with their exponent changes from $\alpha=2$ to $\alpha=1$. 
    $\textbf{(B)}$ Experimental results of the extracted exponent $\alpha$ versus $V_{\oneD}$ (blue circles). 
    The two black dotted lines denote $\alpha=1$ and $\alpha=2$, respectively, while the green dashed line is the theoretical prediction with the effective mass analysis. Error bars are obtained from the linear fit.
    }
    \label{fig:2D_0D}
\end{figure}

When the potential along the tubes $V_\oneD$ increases, two effects occurs. As long as this potential is reasonably small compared to the kinetic energy, its main effect is to modify the effective interactions in the 1D tubes~~\cite{paredes2004tonks}, thus changing the value of the TLL parameter $K$. In this case, the system exhibits a characteristic quantum many-body tunneling behavior, in which interactions within Luttinger liquid play a crucial role, leading to an exponent $\alpha=(4K-1)/2K$ (left panel of Fig.~\ref{fig:2D_0D}A). Then, when this longitudinal potential becomes large, it slices the system into a stack of discrete, disk-shaped layers (right panel of Fig.~\ref{fig:2D_0D}A). The many-body quantum tunneling process is suppressed, since the particles are limited within a 2D plane and the interaction term only serves as an on-site gap. Thus, one would expect the tunneling process reduces to the one-body case.


In the actual experiment, we repeat the $\Jc-T$ measurements and extract the exponent $\alpha$ for various $V_{\oneD}$, with fixed atom number $N=2\times10^5$ in a hexagonal lattice. The results are presented by the blue circles in Fig.~\ref{fig:2D_0D}B. When $V_{\oneD} \le 20 \ \Er$, the system is still governed by the many-body tunneling process between 1D tubes with modified interactions. The validity of 1D description is confirmed by the experimental dimensional crossover diagram as well as the QMC simulations (see details in Supplementary material). 
In this regime, the gradual reduction of $\alpha$ versus $V_{\oneD}$ can be attributed to the increasing 1D interactions. 
The system can be treated as a continuous 1D gas with an equivalent interaction $\gamma^{*}=m^{*}\gamma^{(0)}/m$ where $m^{*}$ is the effective mass and $\gamma^{(0)}$ is the one at $V_{\oneD}=0 \ \Er$.
Correspondingly,  the Luttinger parameter changes into $K^{*}(\gamma^{*})$ and the scaling exponent writes $\alpha^{*}(\gamma^{*})$ (see details in Supplementary material).  In Fig.~\ref{fig:2D_0D}B, we plot the analytical curve $\alpha^{*}=(4K^{*}-1)/2K^{*}$ (green dashed line). It first decreases and saturates at $\alpha=1.5$. The saturation is due to the hard-core limit of $\gamma\gg1$ that gives $\lim_{\gamma\rightarrow\infty}K=1$. The result accounting for the effect of effective mass agrees well with experimental data even up to $V_{\rm 1D}= 30 \ \Er$, suggesting the universality of the scaling law even in the strong interaction regime $K\sim 1$.


As $V_{\oneD}$ continues to increase, $\alpha$ gradually decreases, and saturates to 1 when $V_{\oneD}\geq 60  \ \Er$.
The emergence of $\alpha=1$ can be understood as the crossover to one-body problem. More in details, for sufficiently large $V_{\oneD}$, the tunneling process happens in separated layers. In this regime, the system undergoes a Berezinskii-Kosterlitz-Thouless (BKT) transition, which can be mapped onto the 2D XY model~\cite{Matsubara_1956_2DXYmapping}. The critical temperature writes $T_{\rm c}=0.89J/\kB$~\cite{Kramers_1941_2DXY-Tc,kosterlitz-BKT-1973}.
From a more general point of view, the system essentially undergoes 2D to 0D crossover in this regime and interaction only serves as on-site gap.
Since the only relevant energy scales
are $J$ and $T$, they should follow the scaling $\Jc\sim T$ with the exponent $\alpha=1$.
Thus, we recover the one-body tunneling process in Fig.~\ref{fig:sch}A2. The observed continuous crossover from many-body to one-body tunneling further reveals how interactions renormalize the tunneling dynamics. 

In summary, we measure the universal scaling  of  many-body effects in quantum tunneling process based on a bosonic quantum simulator and provide the first experimental proof for the validity of the quantum field theory prediction for such problems. 
In the weakly-interacting regime, we find a parabolic temperature scaling which is the direct consequence of mean-field approximation, and proof its universality under different conditions. By increasing the longitudinal lattice depth, we find the scaling exponent decrease as the equivalent interaction strength increases, which fits well with the Luttinger liquid prediction. In the extreme case where the longitudinal constrain is large, we find the scaling also reduces to the trivial linear one for one-body problem. 

Our findings significantly advance the understanding of the many-body effect in a quantum tunneling process, bridging the gap between theoretical predictions and experimental observations. 
At the fundamental level, our observation provides insights into the interplay of interaction and temperature during a tunneling process. They allow us to understand how many-body effect renormalize the tunneling. Similar effects are expected for fermions~\cite{quasi-one-dimensional-system-thierry-CR-2004} which opens the door to using the tunneling structure and properties, shaped by interactions and dimensionality as a way to improve phenomena such as superconductivity in mixed dimensionality conditions~\cite{high-Tc-Physica-B-KIVELSON-2002,Field-theory-Lorenzo-PRBL-2026,Improvement-of-superconductivity-Ebot-arxiv-2026}.
Another promising direction is the applicability to tunneling process in other fields. As interaction widely exists in natural systems, whether the universal scaling law detected in atomic systems still holds for stellar or biology system, remains an particularly intriguing open question.


\acknowledgments
The authors thank David Cl\'ement, Tianwei Zhou and Zekai Chen for their helpful discussions. 
\textbf{Funding:} This work is supported by  National Natural Science Foundation of China (Grants No. 92365208), the National Key Research and Development Program of China (Grants No. 2021YFA0718300 and No. 2021YFA1400900) and The Fundamental Research Funds for the Central Universities, Peking University. This work is also supported by the Swiss National Science Foundation under grant number 200020-219400.
\textbf{Author contributions:} The work was conceived by H.S., T.G, H.Y. and X.Z. Experiments were performed by H.S., C.W., and C.Z. Data were analyzed by H.S., C.-K.L., and C.W. Theoretical model and simulation were done by C.-K.L., C.W., and L.P. Experiment preparations were done by C.W., C.Z., and G.S. And X.Z., H.Y. and T.G. supervised this work. H.S., C.-K.L., C.W., T.G. H.Y. and X.Z. wrote the manuscript with input from all authors. All authors discussed the results. 
\textbf{Data Availability:} The data shown in the
main text is available via Zenodo~\cite{shui_zenodo}.

\bibliographystyle{revtex}

\bibliography{my.bib}


 \renewcommand{\theequation}{S\arabic{equation}}
 \setcounter{equation}{0}
 \renewcommand{\thefigure}{S\arabic{figure}}
 \setcounter{figure}{0}
 \renewcommand{\thesection}{S\arabic{section}}
 \setcounter{section}{0}
 \onecolumngrid  
     
 
 \newpage

 {\center \bf \large Supplemental Material for \\}
 {\center \bf \large Universal scaling of many-body effects in quantum tunneling \\ \vspace*{1.cm}
 }

In this supplemental material, we provide details about the experimental details and the effective 1D models with lattice induced interactions.

\section{Experimental details}
\subsection{Control of the atom number and temperature}
\begin{figure}[H]
    \centering
    \includegraphics[width=1\linewidth]{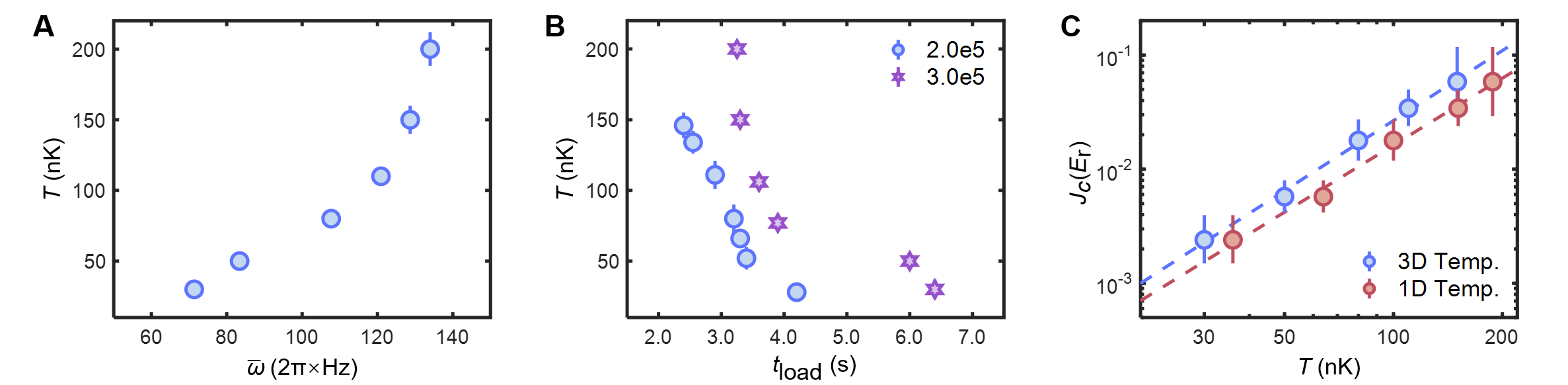}
    \caption{
    \textbf{ Controlling the BEC temperature while fixing the atom number, and the comparison between temperatures before and after loading.
    }
    \textbf{(A)} Final BEC temperatures $T$ of different final geometric mean trap frequencies $\bar{\omega}=\sqrt[3]{\omega_x\omega_y\omega_z}$.    \textbf{(B)} Different BEC temperatures $T$ of different MOT loading times $t_{\rm load}$   with fixed atom numbers $N=2\times10^5$(blue circles) and $N=3\times 10^5$(purple hexagons).
 \textbf{(C)} $\Jc-T$ relations using temperatures at the BEC phase (3D Temperature, blue circles) and the critical point (1D Temperature, red circles) with $V_\oneD = 0 \ \Er$ and $N = 2\times10^5$. The blue and red dashed lines represent linear fits to the corresponding data.  
 Error bars in (A) and (B) represent  the standard deviation of five measurements, while those in (C) are obtained from the piecewise fit.
    }
    \label{fig:S1}
\end{figure}

Our experiment begins with a 3D BEC of $^{87}\text{Rb}$ atoms in the magnetic hyperfine state $\ket{F, m_F} = \ket{1, -1}$. The atoms are first collected into a magneto-optical trap (MOT) and then evaporatively cooled in a crossed optical dipole trap (ODT) to form the BEC. In this work, we need to control the system temperature without changing the atom number. As shown in Fig.~\ref{fig:S1}A, the final temperature of the BEC can be controlled by varying the end trap depth of the ODT during evaporative cooling. Here we use the geometric mean trap frequency $\bar{\omega}=\sqrt[3]{\omega_x\omega_y\omega_z}$ to characterize the trap depth. 
Although the final trap depth also affects the BEC atom number, by adjusting the MOT loading time $t_{\rm load}$, we can precisely control the initial atom number in the MOT and, consequently the final number in the BEC. This method allows us to independently change the BEC temperature while keep the atom number fixed. Fig.~\ref{fig:S1}B shows the loading times of different BEC temperatures and fixed atom numbers.

The loading process of 2D lattices is adiabatic, suggesting the temperature for the coupled 1D system is not the same as the 3D BEC. To confirm that our universal scaling  remains unaffected, we calibrate the temperature of the 1D system. 
Since the system is continuous along $z$ direction, we integrate the momentum distribution over the 2D lattice directions and use the bi-modal fitting to extract the system temperature at $V\lesssim V_c$. We take the group of data with parameters $V_{\oneD}=0 \ \Er$ and $N=2\times10^5$ as an example, and the results are shown in Fig.~\ref{fig:S1}C. We find that all the temperatures of the critical point shift by approximately 30\% compared to the BEC temperatures. Since the scaling factor is decided by the linear fit in the log-log scale, the scaling factor extracted from the 1D temperature is 1.95(10), which is less than $5\%$ different from the 3D temperature.

\subsection{Geometry of the 2D lattice}
\begin{figure}[H]
    \centering
    \includegraphics[width=0.7\linewidth]{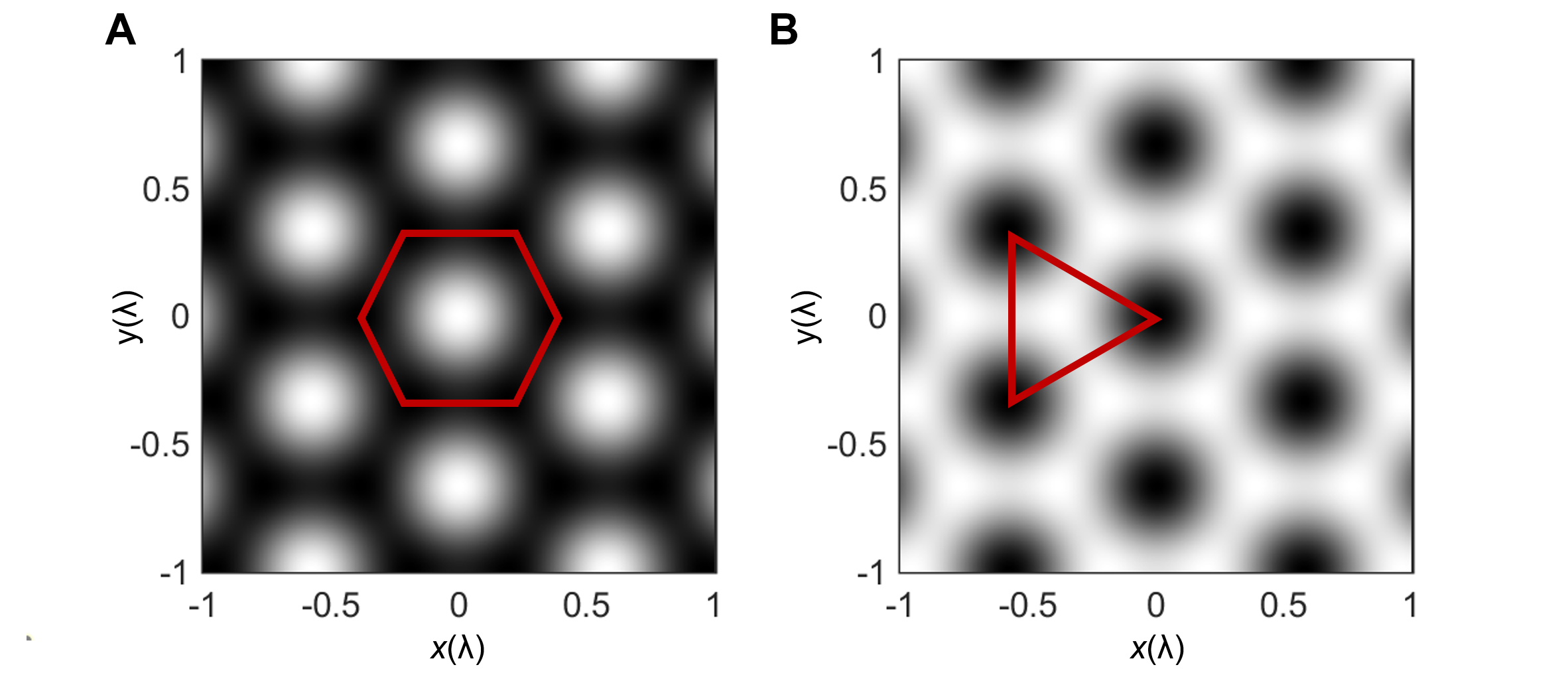}
    \caption{
    \textbf{ Potential distribution of hexagonal and triangular lattice. 
    }
    \textbf{(A)} Hexagonal lattice. The polarizations of the three lattice beams are in the in-plane configuration. Dark and light areas represent the potential valleys and peaks, respectively. The valleys form a hexagonal lattice. \textbf{(B)} Triangular lattice. The polarizations of the three lattice beams are in the out-of-plane configuration. Dark and light areas represent the potential valleys and peaks, respectively. The valleys form a triangular lattice. The red solid lines in both (A) and (B) connect neighboring valleys, highlighting the hexagonal and triangular lattice geometries, respectively.
    }
    \label{fig:S2}
\end{figure}

Our 2D lattice consists of three traveling wave lights with wave vectors $\vec{k_1},\ \vec{k_2},\ \vec{k_3}\quad(k=2\pi/\lambda)$ and same amplitude $E=\frac{\sqrt{V}}{2}$, where $\vec{k_1}=k(\frac{\sqrt{3}}{2},-\frac{1}{2},0), \vec{k_2}=k(\frac{-\sqrt{3}}{2},-\frac{1}{2},0), \vec{k_3}=k(0,1,0)$. All the lights are in the $xy$-plane, and each wave vector separated by $120^\circ$. Their polarization, denoted by $\vec{\epsilon_1},\vec{\epsilon_2}$ and $\vec{\epsilon_3}$, can be switched between in-plane and out-of-plane configurations. In the in-plane case where $\vec{\epsilon_1}=(\frac{1}{2},\frac{\sqrt{3}}{2},0),\vec{\epsilon_2}=(\frac{1}{2},-\frac{\sqrt{3}}{2},0),\vec{\epsilon_3}=(-1,0,0)$, the lattice potential becomes
\begin{align}
    V(x,y,z)&\propto -\left|Ee^{i\vec{k_1}\cdot\vec{r}}\vec{\epsilon_1}+Ee^{i\vec{k_2}\cdot\vec{r}}\vec{\epsilon_2}+Ee^{i\vec{k_3}\cdot\vec{r}}\vec{\epsilon_3}\right|^2 \notag \\ 
    &=-\frac{V}{4}\left(6-2\cos(k\sqrt{3}x)-4\cos(k\frac{\sqrt{3}}{2}x)\cos(k\frac{3}{2}y)\right).
    \label{eqS2}
\end{align}
where the $\vec{r}=(x,y,z)$ and the negative sign comes from the red-detuning of lasers. The valleys of this potential form a hexagonal lattice, as shown inf Fig.~\ref{fig:S2}A.

In the out-of-plane case where $\vec{\epsilon_1}=\vec{\epsilon_2}=\vec{\epsilon_3}=(0,0,1)$, the lattice potential is 
\begin{align}
    V(x,y,z)&\propto -\left|Ee^{i\vec{k_1}\cdot\vec{r}}\vec{\epsilon_1}+Ee^{i\vec{k_2}\cdot\vec{r}}\vec{\epsilon_2}+Ee^{i\vec{k_3}\cdot\vec{r}}\vec{\epsilon_3}\right|^2 \notag \\ 
    &=-\frac{V}{4}\left(3+2\cos(k\sqrt{3}x)+4\cos(k\frac{\sqrt{3}}{2}x)\cos(k\frac{3}{2}y)\right),
    \label{eqS1}
\end{align}
Compared with the in-plane case, the signs of the cosine terms are inverted, which effectively exchanges the peaks and valleys of the potential. Consequently, the resulting lattice exhibits a triangular geometry in the $xy$-plane, as shown in Fig.~\ref{fig:S2}B.

\section{The effective 1D model with lattice induced interactions}
\label{Sec.S2}
As shown in Fig.~\ref{fig:2D_0D}, we present the analytic result for the dependence of the exponent $\alpha$ on the well depth of 1D lattice $V_{\oneD}$. In this section, we establish the range of $V_{\oneD}$ the 1D regime remains valid, and derive $\alpha(V_{\oneD})$ based on the effective mass and the modified Lieb--Liniger parameter.

\subsection{Validity of the 1D regime}

Here, we further provide evidence for the validity of the 1D regime in the range $V_{\oneD}\leq20\ \Er$. From a theoretical perspective, one can examine the superfluid fraction $f_s$ and the single-particle correlation function $g^{(1)}$ along the 1D lattice direction, both of which are accessible via quantum Monte Carlo (QMC) simulations~\cite{PhysRevLett.96.070601,PhysRevE.74.036701}. We employ the path-integral Monte Carlo method following Ref.~\cite{tian2025probinguniversalphasediagram}. In practice, we simulate the Bose--Hubbard model for a homogeneous system to emulate the experimental setup. The simulated lattice geometry consists of a hexagonal lattice in the $xy$-plane and a 1D lattice along the $z$ direction. The simulated system sizes are set to $L_x=20a_x$, $L_y=20a_y$, and $L_z=11a_z$, with $a_j$ the corresponded lattice spacing along $j$ direction. The longitudinal size $L_z$ is chosen to match the experimental setup, while the transverse dimensions $L_x$ and $L_y$ are smaller than those in the experiment. We set the filling to $n=25.5$ per lattice site, corresponding to the central density of a weighted 1D tube~\cite{guo2024observation}. During the simulation, the system is first equilibrated for more than $3\times10^5$ steps to ensure thermalization, followed by an additional $3\times10^5$ steps for statistical sampling.

\begin{figure}[t]
    \centering
    \includegraphics[width=\textwidth]{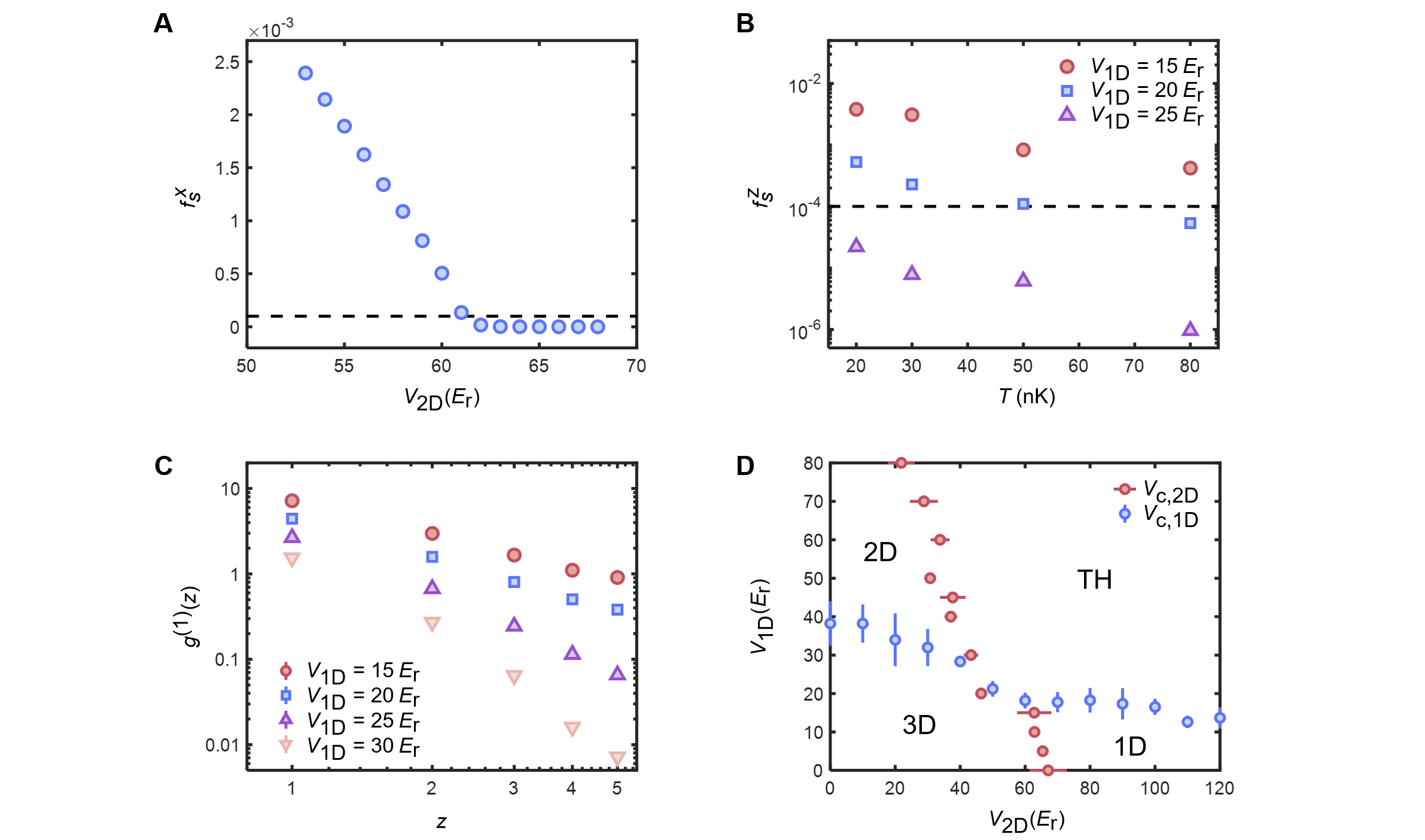}
    \caption{\textbf{Criteria for determining the 1D regime.}
    \textbf{(A)} Determination of the critical lattice depth $\Vc$ from the superfluid fraction along the $x$ direction, $f^x_{s}$, at $T=20\ \mathrm{nK}$ and $V_{\oneD}=20\ \Er$ obtained using QMC simulations. The dashed line indicates the threshold value $1\times10^{-4}$. 
    \textbf{(B)} Superfluid fraction along the $z$ direction, $f^z_{s}$, at $V_{\twoD}=\Vc$ as a function of temperature $T$ for different values of $V_{\oneD}$. The dashed line indicates the same threshold value $1\times10^{-4}$.
    \textbf{(C)} Single-particle correlation function along the $z$ direction, $g^{(1)}(z)$, at $T=80\ \mathrm{nK}$ and $V_{\twoD}=\Vc$ for various values of $V_{\oneD}$.
    \textbf{(D)} Measured phase diagram at $T=30\ \mathrm{nK}$ with $N=2.0\times10^5$ atoms in the experiment. 
    Error bars are obtained from the piecewise fit.
    The QMC results presented in (A)(B)(C) are obtained by simulating a homogeneous system with system sizes $L_x=20a_x$, $L_y=20a_y$, and $L_z=11a_z$. The filling is set to $n=25.5$ per lattice site.
    }
    \label{fig:S3}
\end{figure}

To verify the validity of the 1D regime, we first analyze the superfluid fraction obtained from the QMC simulations. The superfluid fraction along the $i$ direction ($i=x,\ y,\ z$) is defined as $f_s^i = m\kB T\langle W_i^2\rangle / 3\hbar^2\rho L_i$, where $W_i$ is the winding number along the $i$ direction under periodic boundary conditions, $T$ is the temperature, and $\rho$ is the particle density. Since the scaling behavior is governed by the critical tunneling $\Jc\propto T^\alpha$, we examine the validity of the 1D model around the critical point for different values of $V_{\oneD}$. To locate $\Jc$ in the simulations, we first scan $V_{\twoD}$ and calculate $f_s^x$ for given values of $T$ and $V_{\oneD}$. The critical lattice depth $V_{\twoD}=\Vc$ is determined by the criteria $f_s^x < 1\times10^{-4}$, which is approximately the inverse of the simulated system volume. As an example, Fig.~\ref{fig:S3}A shows the calculated $f_s^x$ for $T=20\ \mathrm{nK}$ and $V_{\oneD}=20\ \Er$, yielding $\Vc=61\ \Er$. At the corresponding critical point $\Vc$, we further examine $f_s^z$ for different values of $V_{\oneD}$, which reflects the coherence along the longitudinal direction. The simulation results for $f_s^z$ at $\Vc$ as a function of temperature $T$ are presented in Fig.~\ref{fig:S3}B. We find that $f_s^z$ remains above the threshold value $1\times10^{-4}$ at all calculated temperatures for $V_{\oneD}=15\ \Er$, while it stays around the threshold for $V_{\oneD}=20\ \Er$ and falls below the threshold for $V_{\oneD}=25\ \Er$. These results suggest that the 1D regime is preserved for $V_{\oneD}\leq20\ \Er$.

This criterion for determining the 1D regime can be further verified using the single-particle correlation function $g^{(1)}(z)\equiv g^{(1)}(0,0,z)=\langle \hat{a}^\dagger_j \hat{a}_{j+z}\rangle$, where $z$ denotes the separation along the longitudinal direction. 
In the presence of the longitudinal lattice, the system can still be treated as modulated 1D system if the lattice sites remains coherently coupled, where $g^{(1)}(z)$ exhibits power-law decay correspondingly.
Otherwise, it decays exponentially. In the QMC simulations, we use the worm algorithm and accumulated the samples in open worldline configurations with creation and annihilation move equivalent as
$\langle \hat{a}^\dagger_j \hat{a}_{j+z}\rangle$~\cite{PhysRevLett.96.070601,PhysRevE.74.036701}. Following the same procedure used for analyzing $f_s^z$, we calculate $g^{(1)}(z)$ at the corresponding $\Vc$ for each $V_{\oneD}$ at $T=80\ \mathrm{nK}$, which is the highest temperature used in the experiment. The results are presented in Fig.~\ref{fig:S3}C, showing that $g^{(1)}(z)$ exhibits linear behavior in the log-log plot for $V_{\oneD}\leq20\ \Er$, indicating power-law decay and preserved 1D coherence. In contrast, $g^{(1)}(z)$ deviates from linear behavior when $V_{\oneD}>20\ \Er$. This observation also supports the validity of the 1D regime for $V_{\oneD}\leq20\ \Er$, in agreement with the conclusions obtained from the superfluid fraction.


From the experimental perspective, we can also determine the boundaries of the 1D regime via TOF detection similarly as~\cite{tian2025probinguniversalphasediagram}. By measuring the zero-momentum fraction across a wide range of lattice potentials, we can identify the critical lattice depths governing the dimensional crossovers. Hence the dimensionality of the system under specific parameters can be identified. As an example, the detected phase diagram at $T=30$~nK with $N=2.0\times 10^5$ atoms is shown in Fig.~\ref{fig:S3}D. Clearly, by varying $V_{\mathrm{2D}}$ and $V_{\mathrm{1D}}$ independently, we can identify four regimes, namely the quantum 3D, 2D, 1D regime and the thermal phase. The results reveal that a 3D-1D crossover can happen only for $V_{\mathrm{1D}} \leq 20\,E_{\mathrm{r}}$, which is consistent with theoretical criteria above.


\subsection{Calculation of the effective mass and the modified Lieb--Liniger parameter}

In the following, we derive the relation of exponent $\alpha(V_{\rm 1D})$ and $V_{\rm 1D}$, taking into account the increase of interaction strength with increasing $V_{\rm 1D}$. This relation is obtained under two main approximations.

First, we assume that the 2D lattice $V_{\rm 2D}$ is sufficiently deep to decouple the 1D tubes, leaving the atomic motion restricted to the longitudinal direction of each tube. In this case, the effective mass arises from the 1D lattice. This approximation is justified when $V_{\rm 2D}$ approaches the critical value discussed in the main text. Second, we assume that the 1D gas structure is preserved after imposing the 1D lattice, such that the 1D Hamiltonian in Eq.~\eqref{eq2} remains valid. As demonstrated in Sec.~\ref{Sec.S2}A, this assumption holds for $V_{\oneD}\leq20\ \Er$. Under these approximations, the effective mass induced by the 1D lattice can be incorporated into the interaction parameter, while the mean-field description based on the Luttinger parameter remains applicable.

\begin{figure}[H]
    \centering
    \includegraphics[width=\textwidth]{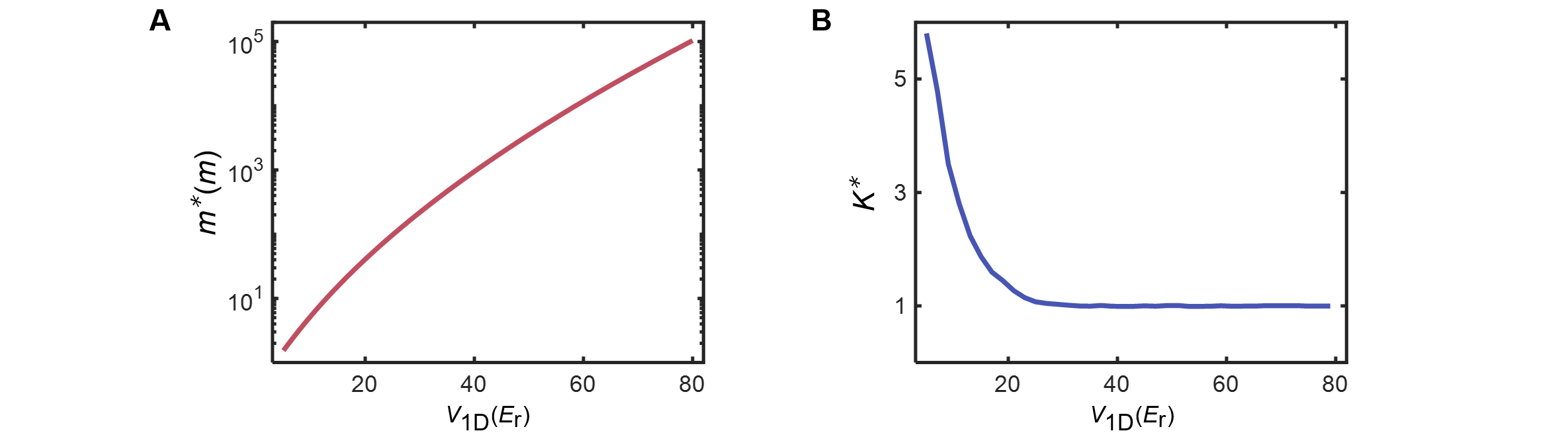}
    \caption{
    \textbf{Calculating the effective mass and Luttinger parameter.}
    \textbf{(A)} Effective mass $m^*$ as a function of the 1D lattice well depth $V_{\oneD}$.
    \textbf{(B)} Corresponding Luttinger parameter $K^*(\gamma^*)$, obtained from the modified Lieb--Liniger parameter $\gamma^*$ that incorporates the effective mass rescaling in Eq.~\eqref{eqS6}.}
    \label{fig:S4}
\end{figure}

In the presence of the 1D lattice, the dispersion relation of the lowest band is
\begin{align}
E(q) = -2J_{\rm 1D}\cos(qa),
\label{eqS3}
\end{align}
where $q$ is the quasi-momentum, $J_{\rm 1D}$ is the tunneling strength along the 1D lattice, and $a$ is the lattice spacing. Expanding the dispersion near $q=0$, we approximate it as
\begin{align}
E(q) \approx E_0 + \frac{\hbar^2 q^2}{2m^*},
\label{eqS4}
\end{align}
where $E_0$ is the energy at the bottom of the band, and $m^*$ is the effective mass. By comparing the two expressions, the ratio between the effective mass and the bare atomic mass is
\begin{align}
\frac{m^*}{m} = \frac{\Er}{\pi^2 J_{\rm 1D}}.
\label{eqS5}
\end{align}
The result of $m^*$ with varying $V_{\oneD}$ is plotted in Fig.~\ref{fig:S4}A. We then introduce a modified Lieb--Liniger parameter $\gamma^*$ defined as
\begin{align}
\gamma^* = \frac{m^*}{m}\gamma^{(0)},
\label{eqS6}
\end{align}
where $\gamma^{(0)}$ is the parameter at $V_{\rm 1D}=0 \ \Er$. The corresponding Luttinger parameter $K^*$ is obtained by mapping $\gamma^*$ onto the Luttinger-liquid theory~\cite{cazalilla2011}. As shown in Fig.~\ref{fig:S4}B, $K^*$ decreases with increasing $V_{\oneD}$, and saturates at the hard-core limit $K^*=1$ when $V_{\oneD}\geq20\ \Er$. Finally, the exponent is given by
\begin{align}
\alpha= \frac{4K^* - 1}{2K^*}.
\label{eqS7}
\end{align}

\end{document}